%% file: Gimondi_CO2_confined_arxiv.tex
\documentclass[
aip,
jcp,%
amsmath,amssymb,
reprint,%
]{revtex4-1}

\usepackage{hyperref}
\pdfoutput=1
\usepackage{graphicx}
\usepackage{amssymb}
\usepackage{amsmath}
\usepackage{blkarray}
\usepackage{multirow}
\usepackage{natbib}
\usepackage{epstopdf}
\usepackage{float}

\usepackage{graphicx}
\usepackage{dcolumn}
\usepackage{bm}

\usepackage{tabu} 

\usepackage{color}
\usepackage{soul}

\begin{document}


\title{CO$_2$ packing polymorphism under confinement in cylindrical nanopores}

\author{Ilaria Gimondi}
\affiliation{Thomas Young Centre and Department of Chemical Engineering, University College London, London WC1E 7JE, UK.}%
\author{Matteo Salvalaglio}%
\email{m.salvalaglio@ucl.ac.uk}
\affiliation{Thomas Young Centre and Department of Chemical Engineering, University College London, London WC1E 7JE, UK.}%

\date{\today}

\begin{abstract}
\input{./tex_files/abstract.tex}
\end{abstract}

\maketitle

\section{\label{sec:intro}Introduction}
\input{./tex_files/intro.tex}

\section{\label{sec:methods}Methods}
\input{./tex_files/methods.tex}
\section{\label{sec:results}Results and Discussion}
\input{./tex_files/results.tex}

\section{\label{sec:conslusions}Conclusions}
\input{./tex_files/conclusion.tex}

\section*{Conflict of interest}
There are no conflicts to declare.

\section*{\label{sec:acknowledgments}Acknowledgements}
\input{./tex_files/acknowledgements.tex}

\section*{Supplementary Material}
See supplementary material for single molecule probability distribution in the model pore, derivation of the ideal random distribution $p_{ref}(\theta)$, details of the setup and results of explorative metadynamics simulations, additional analysis of the radial density profiles. .

\section*{\label{sec:references}References}
\bibliographystyle{unsrt}
\bibliography{./CO2_2}

\end{document}



\title{CO$_2$ packing polymorphism under confinement in cylindrical nanopores: Supplementary Information}

\author{Ilaria Gimondi}
\affiliation{Thomas Young Centre and Department of Chemical Engineering, University College London, London WC1E 7JE, UK.}%
\author{Matteo Salvalaglio}%
\email{m.salvalaglio@ucl.ac.uk}
\affiliation{Thomas Young Centre and Department of Chemical Engineering, University College London, London WC1E 7JE, UK.}%

\date{\today}

\maketitle

\input{./tex_files/SI.tex}


%% file: tex_files/abstract.tex
We investigate the effect of cylindrical nano-confinement on the phase behaviour of a rigid model of carbon dioxide using both molecular dynamics and well tempered metadynamics. To this aim we study a simplified pore model across a parameter space comprising pore diameter, CO$_2$-pore wall potential and CO$_2$ density. In order to systematically identify ordering events within the pore model we devise a generally applicable approach based on the analysis of the distribution of intermolecular orientations. Our simulations suggest that, while confinement in nano-pores inhibits the formation of known crystal structures, it induces a remarkable variety of ordered packings unrelated to their bulk counterparts, and favours the establishment of short range order in the fluid phase. We summarise our findings by proposing a qualitative phase diagram for this model. 

%% file: tex_files/intro.tex
Confinement is known to play a role in the phase behaviour of molecular solids, most notably affecting polymorph selection\cite{polymorph_selection_2004, polymorph_control_2008}. For instance, a paradigmatic example of the dramatic effects of confiment on the spatial arrangement of molecules is provided by water, which, as proven both experimentally and computationally, displays a counter-intuitively complex phase diagram under confinement\cite{Maniwa2002,Striolo2005,Kyakuno2011,Algara-Siller2015,Soper2015,Agrawal2016,Chen2016,Chen2016a}. Understanding polymorphism in confined volumes is relevant both to describe natural processes\cite{Biomineralization2010, Biomineralization2013} as well as for driving rational materials and process development\cite{confined_surfaces_2005,pharmaceuticals_nanoconfinement_2007}. Despite its importance a systematic understanding of confinement effects is still lacking. 

Following up a recent work, in which we have investigated the thermodynamics and mechanism of phase transition between CO$_2$ forms I and III in bulk, we set out to study the effect of confinement on CO$_2$ condensed phases. We tackle this problem by carrying out a systematic analysis of CO$_2$ phase behaviour confined in weakly interacting cylindrical nano pores. Our work has a two-fold aim: on the one hand understanding phase behaviour of confined CO$_2$ is relevant due to its prominent role within the carbon cycle, and the surging needs for mitigating its emissions in atmosphere by implementing capture and storage technologies based on adsorption in porous solids \cite{Benson2005,Orr2009,macdowell2010overview,Krevor2015}. On the other hand due to its modest structural complexity accompanied with a rich phase diagram, CO$_2$ represents a convenient model system to perform extensive sampling of polymorphic transitions at finite temperature\cite{Gimondi2017} and gain insight on general aspects of molecular phase transitions under confinement. 

Both experiments and theory highlight remarkable effects of confinements on CO$_2$ phase behavior. For instance, the density of confined CO$_2$ can significantly exceed the fluid phase bulk density \cite{Papadopoulos2001,Steriotis2004,Melnichenko2010,Rother2012,Sanghi2012,Elola2016} reaching values comparable to that of solid CO$_2$ and displaying signs of orientational correlations and structural rearrangements\cite{Steriotis2004,Papadopoulos2001}.
 
Several examples of molecular modelling studies can be found investigating the transport of CO$_2$ in a confined fluid phase \cite{SlitClayPore2005,CO2inClayH2010,Sanghi2012}, and the modelling literature characterising structural features of CO$_2$ under confinement is limited. 
An insightful work from Elola and Rodriguez has recently proposed a detailed analysis of the CO$_2$ fluid structure within silica pores, highlighting how confinement induces a layered arrangement in the fluid phase and determines a significant slowdown of both translational and rotational diffusion of CO$_2$ molecules\cite{Elola2016}.

In our work we aim complementing the state of the art by understanding whether confinement promotes or inhibits the formation of ordered phases.
To this aim we systematically analyse the intermolecular structural organisation of CO$_2$ confined in a simplified pore geometry consisting of a cylinder of Lennard-Jones (LJ) particles. We carry out a systematic study as a function of the cylinder radius (spanning from 1 to 5 nm), the nominal density of CO$_2$ molecules (5 - 15 molecules/nm$^3$), and the interaction potential with the pore walls.
We investigate this parameter space by systematically carrying out unbiased molecular dynamics simulations starting from liquid-like initial conditions, and by enhancing the exploration of ordered molecular packings with metadynamics.  

Our analysis shows that, while the formation of known solid CO$_2$ arrangements is inhibited, confinement unveils a rather complex behaviour. 
Depending on the location in parameter space, confined CO$_2$ can either approach a completely disordered fluid state, exhibit short-range orientation correlations or arrange into long-range ordered packings. By systematically analysing relative orientations distributions we have systematically detected ordering events from molecular trajectories and classified ordered configurations.

This paper is organized as follows: firstly, in section 2, we introduce the model system, methods, parameters space, and the analysis tools developed to characterise ordering transitions. In section 3 we present and discuss our results, and finally in section 4 we summarise our findings by proposing a qualitative phase diagram in the space defined the nominal density of the pore and its normalised radius. 

%% file: tex_files/methods.tex
\begin{figure*}[t]
\includegraphics[width=1\textwidth]{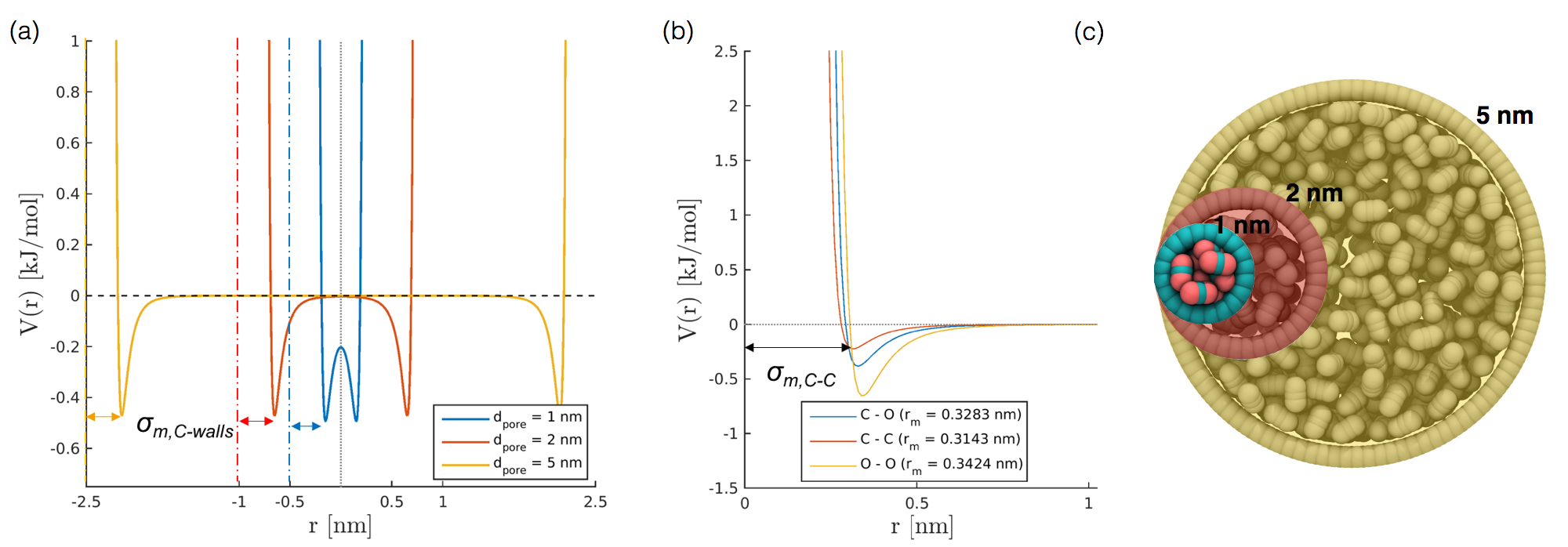}
\caption{Simulations setup. (a) Representation of the LJ potential for the interaction between the pore wall and the carbon atom of CO$_2$ we report three pore sizes (d$_{pore}$ = 1, 2, 5 nm), for which the zero on the $x$ axis represents the cylinder axis and the walls are located at $\pm$0.5, $\pm$1 and $\pm$2.5 nm, respectively; $\sigma_{wall}$ is set to 0.34 nm. Arrows highlight $\sigma_{m,C-wall}$ (b) Intermolecular LJ potential characterising CC, CO and OO non bonded interactions. The $x$ axis represents the interatomic distance; the black  arrow indicates $\sigma_{m,C-C}$. (c) Representation of the cross-section of the initial configuration for simulations with $d_{pore}$= 1, 2, and 5 nm.}
\label{imageLJ}
\end{figure*}

To investigate the effect of confinement on liquid carbon dioxide we have employed both standard molecular dynamics (MD)\cite{Tuckerman2010,Frenkel2002,Palmer2015} and well-tempered metadynamics (WTMetaD)\cite{Barducci2008}. For a detailed description of metadynamics we refer to Barducci et al.\cite{Barducci2008,Barducci2011}, and Valsson et al.\cite{Valsson2016}, and for a brief overview of its applications in crystallisation studies to Giberti et al.\cite{Giberti2015}. In the following we report the setup of the molecular model investigated, the details of MD simulations and WTmetaD simulations setup, and the analysis approach implemented to detect and characterise the emergence of an ordered phase from an initial liquid state. 

\paragraph*{CO$_2$ Potential.} In order to model CO$_2$ we employ the Transferable Potentials for Phase Equilibria (TraPPE) force field\cite{POTOFF1999,Potoff2001} (Table~\ref{table_TraPPE}) with the introduction of dummy atoms to maintain molecules rigid with a 180$^\circ$ angle without inducing instability\cite{Sanghi2012}. This choice is consistent with our previous work on carbon dioxide polymorphism at high pressure \cite{Gimondi2017}, and allows for a direct comparison with results obtained in bulk.
\begin{table}[h]
\centering
\setlength\extrarowheight{2pt}
\caption{Parameters for the TraPPE force field}
\begin{tabular}{ ccccc }
 \hline
m$_C$ & m$_O$ & $\sigma_{C-C}$ & $\sigma_{O-O}$& $\epsilon_{C-C}$ \\ 
$ $ & $ $ & \small{[nm]} & \small{[nm]} & \small{[kJ/mol]} \\ 
 \hline
 12 & 16 & 0.280 & 0.305 & 0.224 \\[0.5ex] 
\hline \hline
$\epsilon_{O-O}$  &  q$_c$  & q$_o$  & $l_{C-O}$& $\alpha_{O-C-O}$ \\
\small{ [kJ/mol]}& \small{[e]}& \small{[e]} & \small{ [\AA] } & \small{[$^{\circ}$]} \\ 
\hline
0.657 &  0.70 & -0.35 & 1.160 & 180\\
\hline
\end{tabular}
\label{table_TraPPE}
\end{table}

\paragraph*{Pore Model.} To develop a simplified model to study confinement in a weakly interacting porous medium, we build a single-wall cylindrical nanopore of LJ particles. We consider different pore diameters (1, 1.3, 2, 5 nm) with a constant wall density of 33.104 atoms/nm$^2$ and a height of 10 nm for all diameters except the case with diameter 5 nm, where the height has been limited to 5 nm as well for the sake of computational efficiency. The cylinder axis aligns with \textit{z}. We also investigate the effect induced by variations in the wall potential induced by keeping constant $\epsilon_{wall}$ while varying the $\sigma_{wall}$, in the interval between 0.253 to 0.405 nm. 

\paragraph*{Simulation set up.} For our simulations we choose temperature conditions of 323 K and nominal carbon dioxide densities compatible with supercritical CO$_2$\cite{Krevor2015,Melnichenko2010,Sanghi2012}. 
Firstly, we run NVT molecular dynamics simulations of the described system, with Bussi-Donadio-Parrinello\cite{Bussi2007} thermostat. We apply periodic boundary conditions (\textit{pbc}) to ensure continuity along  \textit{z}, i.e. the cylinder axis. To avoid overlapping effects in the radial direction from periodic images, the simulation box is much larger (around 10 times) than the pore diameter. Lennard-Jones potential with Lorentz-Berthelot combination rules and long-range corrections is employed. Typical MD simulations are 20 ns-long, starting from disordered configurations equilibrated for 100 ps.

We use the same setup to carry out WTMetaD simulations aimed at exploring the space of accessible packings beyond the timescale limit of unbiased MD. As collective variable we employ an order parameter, hereafter indicated with $\lambda$, that expresses the degree of crystallinity of a system based on the local environment around each molecule. A complete description of the mathematical formulation of $\lambda$  has been reported by Giberti et al.\cite{Giberti2015_2}. Explorative WTMetaD simuations are carried out biasing two distinct formulations of the $\lambda$ order parameter: $\lambda_I$, tuned to capture the molecular arrangement of CO$_2$ form I, which we recently developed to investigate CO$_2$ polymorphism at high pressure\cite{Gimondi2017}; and $\lambda_B$, based instead on the characteristic angles and local density of the most abundant ordered structure observed under confinement (configuration B in Table~\ref{imageordered}). For details on the parameters used to define both $\lambda_I$ and $\lambda_B$ refer to Table~\ref{table_lambda}. 
Further details about the WTMetaD set up are reported in the Supporting Information.

\begin{table}[h!]
\centering
\setlength\extrarowheight{5pt}
\caption{Tuning of the $\lambda$-order parameters.  The table reports $\theta_{1}$, its supplementary $\theta_{2}$, the associated width of the Gaussian, $\delta$, which for symmetry reasons is the same for both angles.  The cut-off values for the number of neighbours and the coordination shell are presented as well.}
\begin{tabular}{ c|ccccc } 
 \hline
  & $\theta_{1}$ [$^{\circ}$] & $\theta_{2}$ [$^{\circ}$] & $\delta_{1}=\delta_{2}$ [$^{\circ}$] & n$_{cut}$ [-] & r$_{cut}$ [\AA] \\
  \hline \hline
   $\lambda_I$ & 70.47 & 108.86 & 14.32 & 5 & 4 \\
   $\lambda_B$ & 46.18 & 133.82 & 5.21 & 3 & 4 \\
\hline  
\end{tabular}
\label{table_lambda}
\end{table}

\paragraph*{Parameter space representation.} In order to represent our results throughout parameter space in a rational and efficient form we define a single adimensional radius, $r^\prime$, which accounts for variations in both LJ potential ($\sigma_{wall}$) and pore size and quantifies the void space available within the pore\cite{Kyakuno2011}. The adimensional radius $r^\prime$ is defined by the following expression: 
\begin{equation}
r^\prime = \frac{r_{pore}-\sigma_{m,C-wall}}{\sigma_{m,C-C}}
\label{Gamma}
\end{equation}
where $\sigma_{m,C-wall}$ represents the distance from the wall to the minimum of the C-wall potential (as in Figure~\ref{imageLJ}(a), for $\sigma_{wall}$ = 0.34 nm) and $\sigma_{m,C-C}$ is the position of the minimum of the C-C interaction. Values of $r^\prime$ for all simulations performed in this work are reported in Table~\ref{table_rprime}. 
The second parameter is the nominal density of CO$_2$ molecules within the pore $\rho_{CO_2}$, computed accounting for the volume of the cylinder of diameter $d_{pore}$.

\begin{figure}[b]
\includegraphics[width=0.5\textwidth]{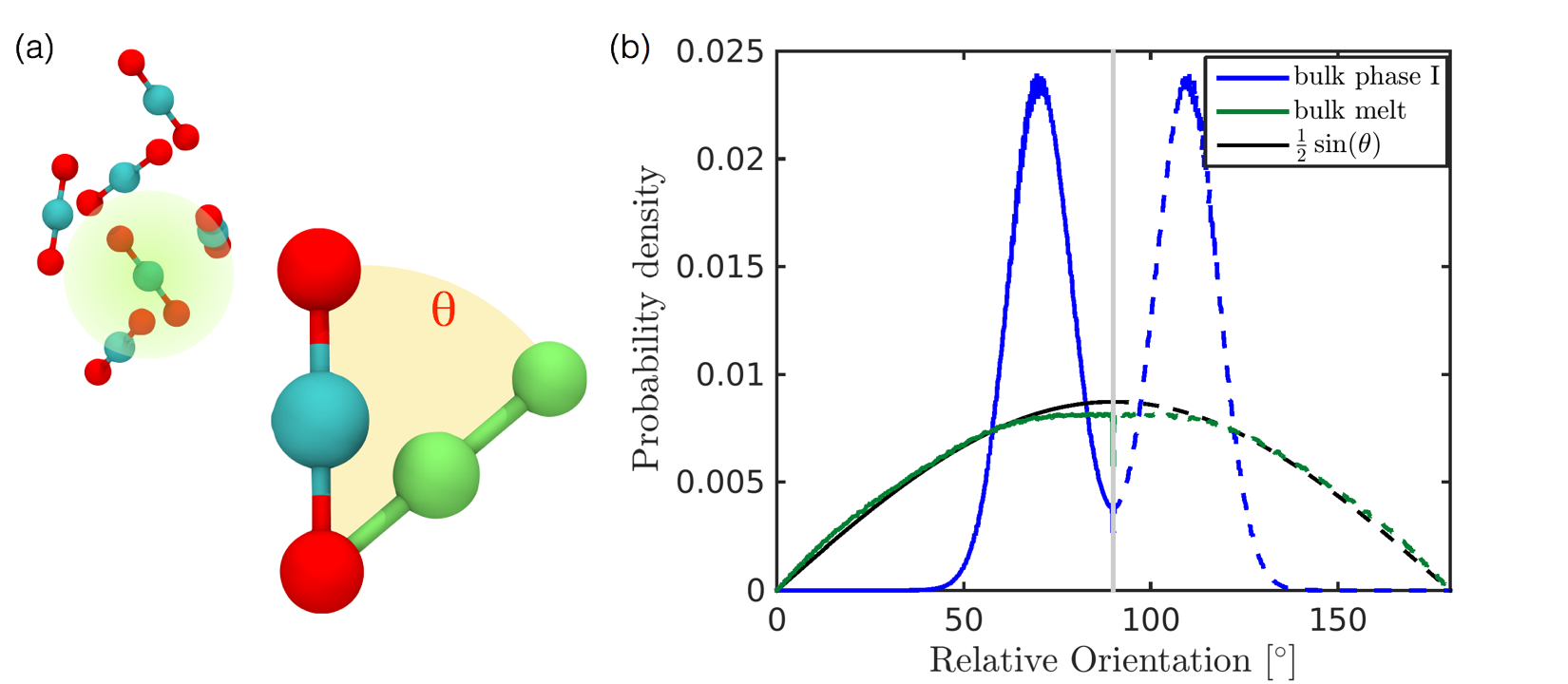}
\caption{(a) The distribution of relative orientations is built considering all the angles between an assigned molecule and its nearest neighbours, falling within a sphere of radius $r_{cut}$ represented in transparent green. The relative orientation is computed as the elevation angle between the axis of two CO$_2$ molecules. (b) Characteristic angle distribution for bulk phases, in particular melt (green) and phase I (blue); the analytical expression for the random distribution in melt is also reported in black.}
\label{imagebulk}
\end{figure}

\begin{figure*}[t]
\includegraphics[width=1.0\textwidth]{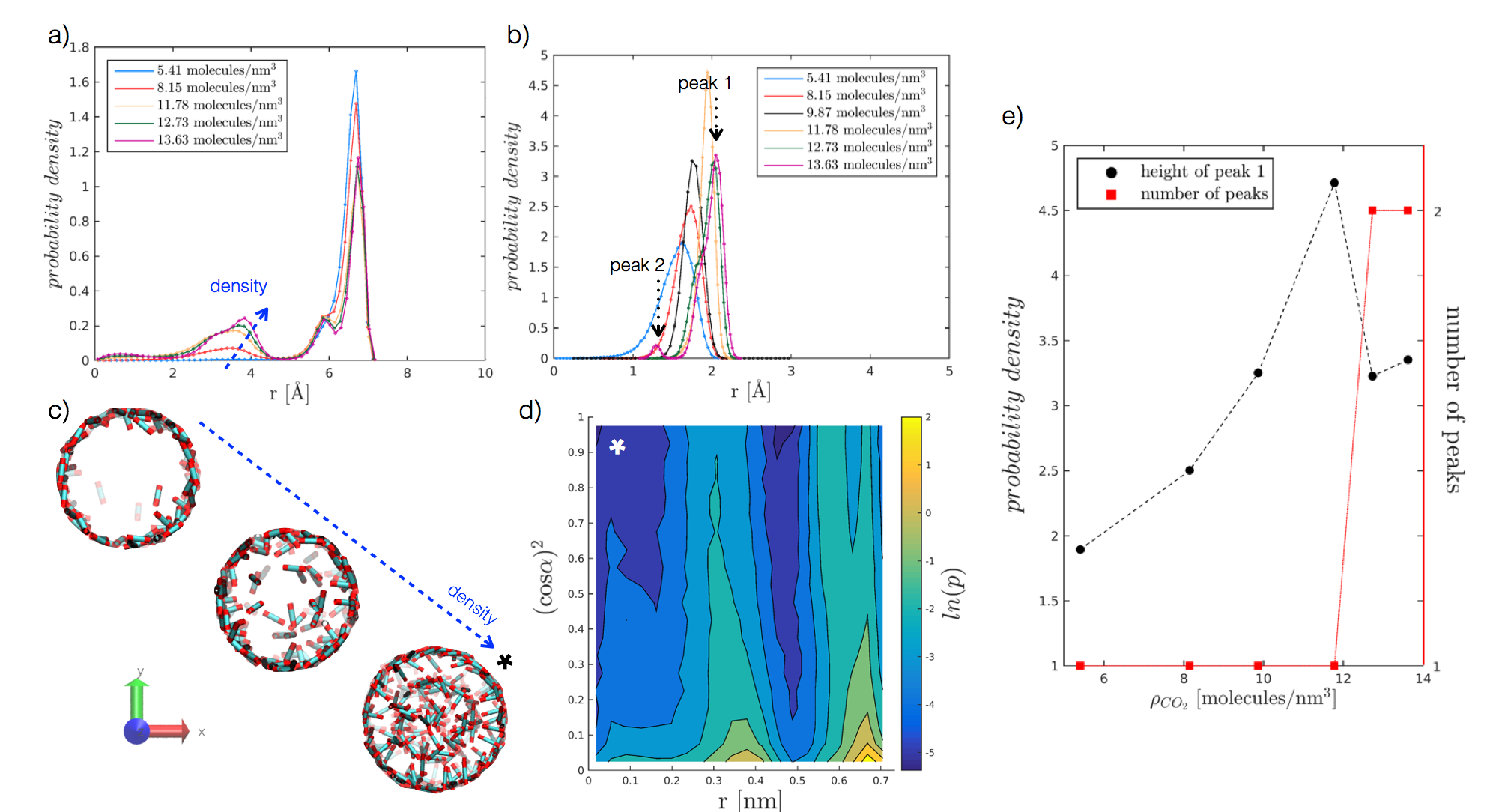}
\caption{Probability density profiles associated with the position of CO$_2$ molecules in the radial direction of the pore, for the range of densities investigated at different values of $r^\prime$. In particular, (a) refers to $r^\prime$ = 2.075, and (b) to $r^\prime$ = 0.484. Along the abscissa, i.e. the pore radius, zero corresponds to the cylinder axis, while the maximum value reported to the radius of the pore. For the case in (b), (e) further analyses the height  of the adsorbed layer peak and the number of peaks as a function of $\rho_{CO_2}$. For the case in (a), instead, (c) reports snapshots of hollow to filled structures for increasing $\rho_{CO_2}$, from 5.41 to 8.15 to 12.73 molecules/nm$^3$. More in detail, (d) presents the probability density of the angle $\alpha$ as a function of the radial distance, for $r^\prime$ = 2.075 and $\rho_{CO_2}$ = 12.73 molecules/nm$^3$. The angle $\alpha$, reported as (cos$\alpha$)$^2$, is evaluated between CO$_2$ molecular axis and the vector normal to the pore walls; to enhance the differences on the plot the probability density is represented by its logarithm. The $x$-axis convention is the same as before, with cylinder axis in zero.}
\label{imagepeakFES}
\end{figure*}

\begin{table}[htbp]
\centering
\setlength\extrarowheight{2pt}
\caption{Value of $r^\prime$ for the combinations of d$_{pore}$ and $\sigma_{wall}$ investigated. }
\begin{tabular}{cp{1cm}|p{1cm}p{1cm}p{1cm}p{1cm}}
\hline
\multicolumn{2}{c|}{\multirow{2}{*}{$r^\prime$}} & \multicolumn{4}{c}{d$_{pore}$ [nm]}  \\ 
\multicolumn{2}{c|}{} & 1 & 1.3 & 2 & 5 \\ \hline \hline
\multirow{ 5}{*}{$\sigma_{wall}$ [nm]} 
& 0.253	& 0.639	& 1.116	& 2.230	& 7.003 \\
& 0.315	& 0.528	& 1.006	& 2.119	& 6.892 \\
& 0.340	& 0.484	& 0.961	& 2.075	& 6.847 \\
& 0.372	& 0.427	& 0.904	& 2.017	& 6.790 \\
& 0.405	& 0.368	& 0.845	& 1.959	& 6.731 \\ \hline
\end{tabular}
\label{table_rprime}
\end{table}

\paragraph*{Systematic detection of ordered arrangements.} In the definition of ordered parameters such as $\lambda$ the identification of characteristic relative orientations is key\cite{Salvalaglio2012,Giberti2015,Gimondi2017}. This is typically based on the analysis of the distribution of relative orientations that characterises a relevant structure, as extensively discussed for bulk CO$_2$ in Ref. \cite{Gimondi2017}. Indeed the relative orientation between molecular axes of neighbouring molecules represents a \textit{fingerprint} of each arrangement and in principle allows to distinguish not only between liquid and solid, but also among different solid structures. An example of such distributions for bulk dry ice and liquid CO$_2$ is reported in Figure~\ref{imagebulk}. In the following we build on this observation to systematically detect ordering phenomena in within molecular trajectories.

As reported in Figure~\ref{imagebulk}, crystal and liquid carbon dioxide display significantly different characteristic orientation distributions. In order to quantitatively capture such difference we apply the Bhattacharyya distance\cite{Bhattacharyya1943}, a metric that quantifies the \emph{dis}similarity of two probability densities. The Bhattacharyya distance D$_B$ between two angle distributions $p\left(\theta\right)$ and $q\left(\theta\right)$ is defined as: 
\begin{equation}
D_B\left(p,q\right) = -\ln\left(C_B\left(p,q\right)\right)
\label{Bhatta_D}
\end{equation}
where $C_B$ is the Bhattacharyya coefficient, which measures the overlap between $p\left(\theta\right)$ and $q\left(\theta\right)$ defined as: 
\begin{equation}
C_B\left(p,q\right) = \int \sqrt{p\left(\theta\right)q\left(\theta\right)}d\theta
\label{Bhatta_C}
\end{equation}
In our analysis we take advantage of the fact that a completely random arrangement of linear molecules displays a relative orientation probability density characterised by the functional form $p_{ref}=1/2\sin\theta$. For a derivation of the reference distribution functional form see the Supporting Information. As shown in Figure ~\ref{imagebulk} the distribution of angles in the bulk of liquid CO$_2$ closely resembles the theoretical random distribution $p_{ref}$. 

Given a molecular trajectory, we can therefore define a time dependent measure of the deviation from an ideal random packing of CO$_2$ molecules as: 
\begin{equation}
D_B\left(t\right) = -\ln\int{\sqrt{\frac{1}{2}p\left(\theta,t\right)\sin{\theta}}d\theta}
\label{Bhatta_D}
\end{equation}
where $p(\theta,t)$ is the instantaneous probability density of relative orientations.
 
The construction of $p(\theta,t)$ is straightforward for carbon dioxide, due to its linear, symmetrical geometry. For each molecule we evaluate the polar angle between its molecular axis and the analogous vector in each one of its nearest neighbours, i.e.  molecules within a cut-off radius of 4 \AA{} for ordered phases and 5 \AA{} for the liquid as shown in Figure \ref{imagebulk}(a). 

\paragraph*{Tools.} Pores are built using an in-house FORTRAN code, while we use Packmol\cite{Truhlar2009} to insert carbon dioxide molecules at random positions and orientations in order to avoid any creation of patterns. For MD and WTMetaD simulations we employ Gromacs 5.2.1\cite{Abraham2015} and Plumed 2.2\cite{Tribello2014}. Post-processing of the results is carried on with MATLAB (R2015a), Visual Molecular Dynamics (VMD)\cite{HUMP96} and Plumed 2.2\cite{Tribello2014}.

%% file: tex_files/results.tex
\begin{figure*}[t]
\includegraphics[width=1\textwidth]{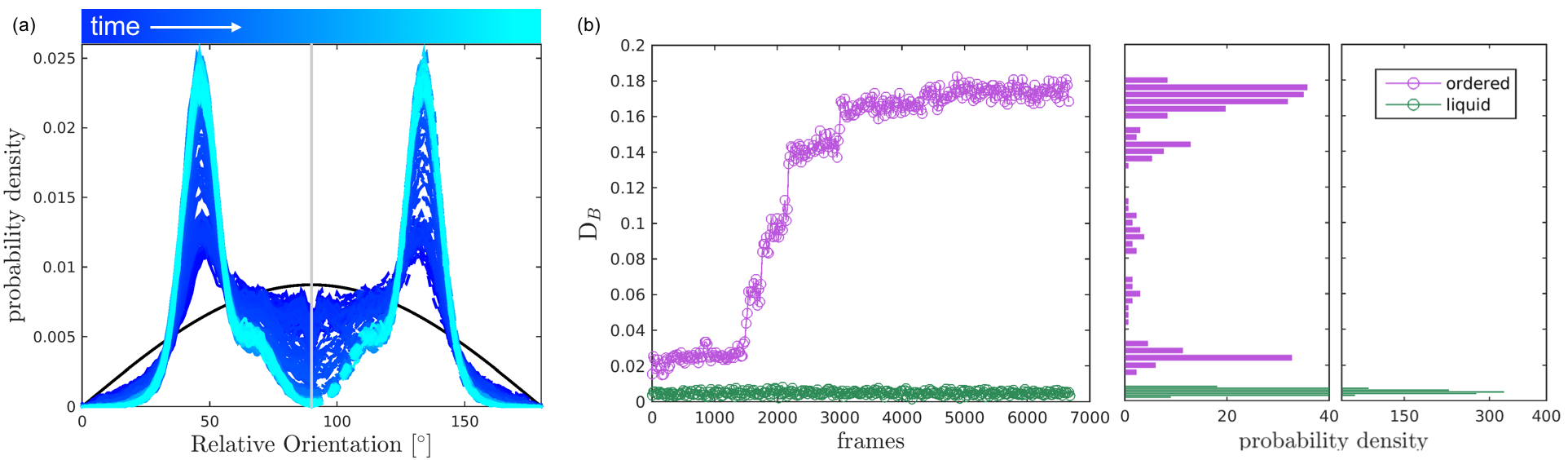}
\caption{Bhattacharyya analysis of an ordering phenomenon. (a) Temporal evolution of the angle distribution, $p(\theta,t)$, for $r^\prime$ = 0.484 - $\rho_{CO_2}$ = 11.78 molecules/nm$^3$, where colour dark blue to light blue represents time increment, while the black distribution is the probability density for melt, employed as reference $p_{ref}$ for evaluating D$_B$. (b) Bhattacharyya distance, D$_B$, as a function of time for an example of ordering (violet, $r^\prime$-$\rho_{CO_2}$ as (a)) and fluid (green, $r^\prime$ =  0.961 - $\rho_{CO_2}$ = 5.41 molecules/nm$^3$ conditions; the probability density of D$_B$ is also reported in the form of a histogram.}
\label{imageBhatta}
\end{figure*}

In this section we present and discuss the results of our molecular dynamics investigation of the effect of confinement on CO$_2$ ordering.  Preliminary simulations were carried out to investigate the equilibrium distribution of a single CO$_2$ molecule in the presence of a cylindrical LJ wall and are discussed in Supporting Information. In the following we begin by discussing the arrangement of CO$_2$ molecules within the pore as a function of pore size and density, and we continue by reporting an analysis of the ordered arrangements identified within our parameter space. 

\paragraph*{Molecular arrangement within the pore.} We begin by analysing the arrangement of CO$_2$ molecules in the pores at increasing $\rho_{CO_2}$. In the case of low densities we observe the formation of a single layer of adsorbed CO$_2$ molecules in contact with the pore surface associated with a hollow structure. With increasing $\rho_{CO_2}$ larger pores (d$_{pore}$ = 2 and 5 nm), can accommodate a second layer of CO$_2$, and then display bulk-like filling as shown in Figure~\ref{imagepeakFES}(a,c).  This finding is in agreement with the results obtained by Elola and Rodriguez\cite{Elola2016} for CO$_2$ confined in silica pores.

Furthermore, we analyse CO$_2$ molecular orientation within the pore by evaluating the angle $\alpha$ between the molecular axis and the normal to the pore surface. As shown in Figure~\ref{imagepeakFES} (d), we can clearly note the high probability regions in the r,$\left(\cos\alpha\right)^2$ plane, corresponding to individual layers. We note that in both layers $\alpha$ is narrowly distributed around 90$^\circ$. This indicates that CO$_2$ molecules arrange parallel to the pore surface, in agreement with observations reported in the literature\cite{Vishnyakov1999,Steriotis2004,Elola2016,Sanghi2012}. 

In smaller pores (d$_{pore}$ = 1 and 1.3 nm) a layered structure is not observed and the decrease in height of the dominant peak (peak 1 in Figure~\ref{imagepeakFES}(b)), accompanied by the appearance of a secondary peak (peak 2 in Figure~\ref{imagepeakFES}(b)) is instead associated with a reorganisation of the CO$_2$ packing within the pore (see Figure~\ref{imagepeakFES}(b,e)). For instance, we note that at d$_{pore}$=1.3 nm, CO$_2$ arranges as a chain along the \textit{z} axis for densities equal or above 9.87 molecules/nm$^3$.

\paragraph*{Detecting emerging order.} After discussing the position and alignment of CO$_2$ molecules with respect to the pore wall, we delve into the analysis of intermolecular arrangement. 
In particular, we analyse the arrangement of the local environment around each CO$_2$ molecule by monitoring the distribution of relative orientations. As mentioned in section 2, this information constitutes a structural \textit{fingerprint} and can discriminates not only between ordered and disordered arrangements but also among different ordered packings. 

As discussed in section 2, to detect the unfolding of ordering events we compute the Bhattacharyya distance, D$_B(t)$ with respect to an ideal random packing for each MD trajectory.  When ordered arrangements nucleate within a pore, the D$_B$ metric displays a sharp increase. An example of this behaviour in a typical trajectory is reported in Figure~\ref{imageBhatta}, where in (b) we compare the trend of D$_B$(t) for a trajectory undergoing an ordering transition (violet) and a trajectory that remains stable in the fluid state (green). In Figure~\ref{imageBhatta}(a) the evolution of the instantaneous intermolecular angle distribution $p(\theta,t)$ is reported together with the $p_{ref}(\theta)$ corresponding to an ideal random packing.
\begin{figure*}[t]
\includegraphics[width=1\textwidth]{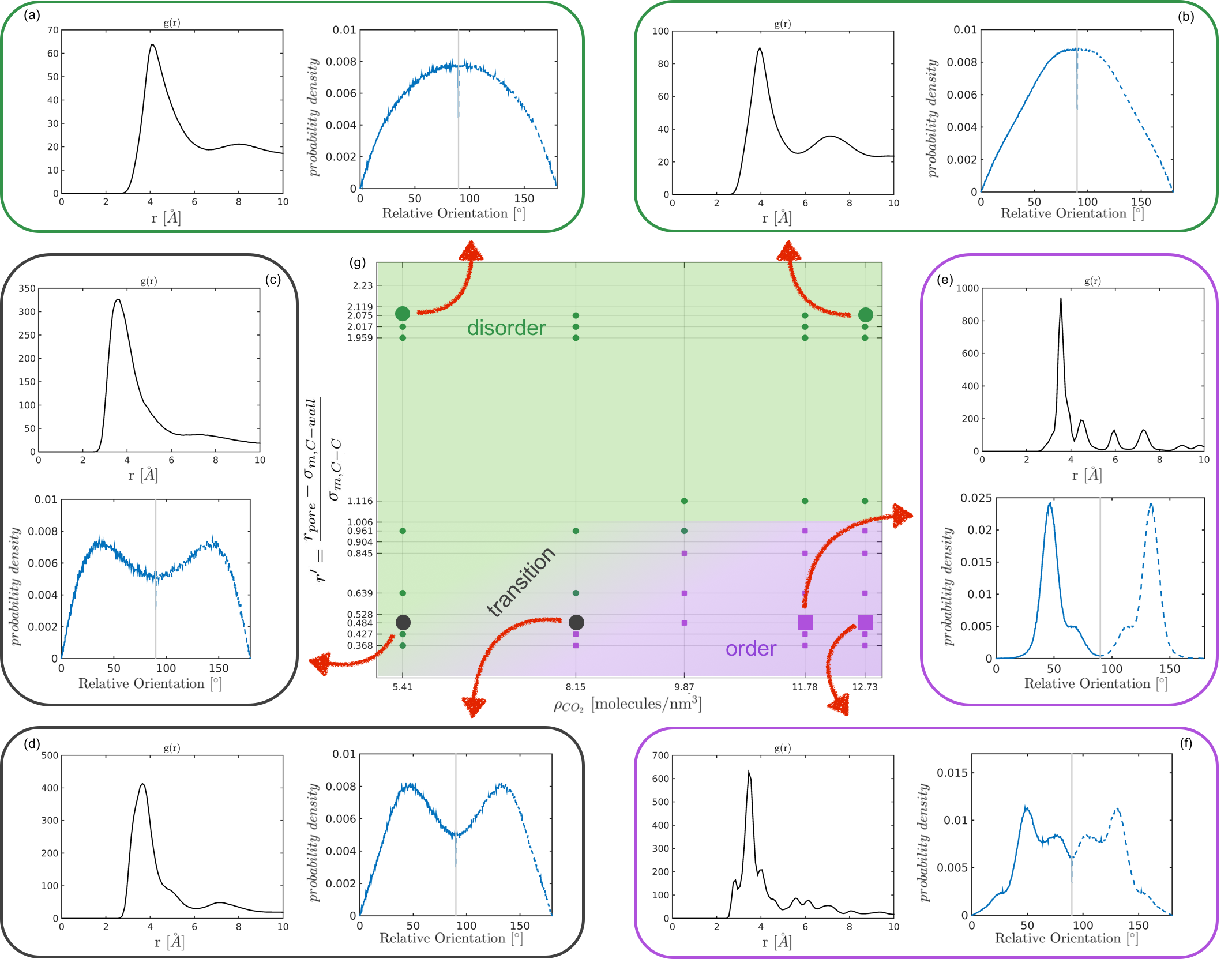}
\caption{Radial pair distribution function, \textit{g(r)} (black) and angle distribution (blue) for illustrative cases of liquid (a-b), transition (c-d), and ordered structures (e-f). In particular, (a) refers to $r^\prime$ = 2.075 - $\rho_{CO_2}$ = 5.41 molecules/nm$^3$, (b) to $r^\prime$ = 2.075 - $\rho_{CO_2}$ = 12.73 molecules/nm$^3$, (c) to $r^\prime$ = 0.484 - $\rho_{CO_2}$ = 5.41 molecules/nm$^3$, (d) to $r^\prime$ = 0.484 - $\rho_{CO_2}$ = 8.15 molecules/nm$^3$, (e) to $r^\prime$ = 0.484 - $\rho_{CO_2}$ = 11.78 molecules/nm$^3$, and (f) to $r^\prime$ = 0.484 - $\rho_{CO_2}$ = 12.73 molecules/nm$^3$. Green areas on the phase diagram refer to fluid state, violet to ordered, while violet-to-green colour to transition structures}
\label{imagegofr}
\end{figure*}

\paragraph*{Mapping phase behaviour in the $(\rho_{CO_2},r^\prime)$ parameter space.} The systematic analysis of the deviation from an ideal random packing and the contextual detection of the appearance of ordered phases from a liquid configuration allows to clearly map the qualitative behaviour of confined CO$_2$ within the $(\rho_{CO_2},r^\prime)$ parameter space. In particular we can identify areas where CO$_2$ under confinement spontaneously evolves to ordered packings within the characteristic timescale of in MD simulations.
To further analyse our finding, we complement the analysis of the angle distribution by reporting the C-C radial pair distribution function \textit{g(r)}. Inspecting both \textit{g(r)} and relative orientation distribution we can identify three regions in parameter space: a disordered region (Figure~\ref{imagegofr} green area, examples (a-b)), an ordered region (Figure~\ref{imagegofr} violet area, examples (e-f)), and a transition region between them (Figure~\ref{imagegofr} green-to-violet area, examples (c-d)). 
We note that in the disordered region, at large $r^\prime$ and low density, the distribution of relative orientations approaches the ideal random distribution and the \textit{g(r)} displays the short-range order typical of a liquid. On the contrary, in the ordered region both the angle distribution and the \textit{g(r)}, deviate from the typical liquid behaviour denoting long-range order through the presence of characteristic peaks in both orientation distribution and \textit{g(r)}. 
Finally, the transition region denotes a peculiar behaviour. In this region of parameter space, while \textit{g(r)} maintains a close resemblance to that of a liquid phase, the distribution of relative orientations displays significant deviations from an ideal random packing suggesting the development of correlation in the relative orientation of CO$_2$ molecules. 

From the analysis of the transitions obtained in the structured region of parameter space we notice a finer grained level of complexity emerging. As displayed by the qualitative comparison of both the distribution of relative orientations and the \textit{g(r)} reported in examples Figure~\ref{imagegofr}(e) and (f) multiple ordered packings can be obtained.

\paragraph*{Stable ordered packings.} We now move on to present a detailed analysis of the ordered structures encountered within the ordered region of the $(\rho_{CO_2},r^\prime)$ phase diagram. We shall note that, in order to confirm the finding emerging from unbiased simulations, we have applied WTMetaD simulations to further sample the configuration space of confined CO$_2$. This allows to verify that in the disordered or transition regions ordered arrangements cannot be found or, when found do not states with a finite lifetime and spontaneously revert to liquid in unbiased MD.
\newline Nevertheless, in the region of the phase diagram where confinement-induced order emerges, we identify four different stable arrangements. We assign to such arrangements a label progressing from A to D, based on their structural features. As we can see in Table~\ref{imageordered}, progressing from A to D these structures present an increasing number of molecules in their repeating unit. Moreover, the same classification applies when analysing the geometry of the apparent ring that characterises the horizontal cross-section of the pore (see Table~\ref{imageordered} ). The number of ring members is 3 in A, 4 in B, and 6 in both C and D, with D displaying one molecule in the middle of the pore oriented parallel to the pore longitudinal axis. 

\begin{table*}[t]
\caption{Molecular representation of the arrangement and characteristic angles distribution for the ordered structures identified. Such structures are named A to D according to the apparent number of ring members when observed from the top. The angles distributions are obtained at $\Gamma$ = 0.484 - $\rho_{CO_2}$ = 9.87, at $r^\prime$ = 0.484 - $\rho_{CO_2}$ = 11.78, at $r^\prime$ = 0.484 - $\rho_{CO_2}$ = 12.73, at $r^\prime$ = 0.961 - $\rho_{CO_2}$ = 12.73, for A, B, C, and D, respectively.}
\begin{center}
\begin{tabular}{ |c| c | c | }
\hline 
\rule[-1ex]{0pt}{2.5ex}  Label & Snaphots & Relative Orientation Distribution \\ 
\hline 
\hline 
\rule[-1ex]{0pt}{2.5ex}  A & \includegraphics[width=0.35\textwidth]{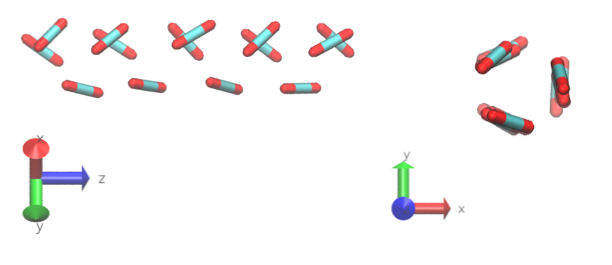} & \includegraphics[width=0.20\textwidth]{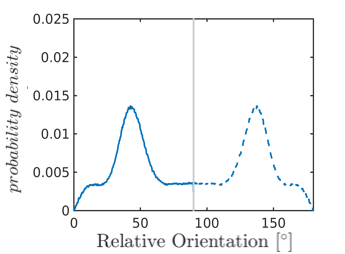} \\ 
\hline 
\rule[-1ex]{0pt}{2.5ex}  B & \includegraphics[width=0.35\textwidth]{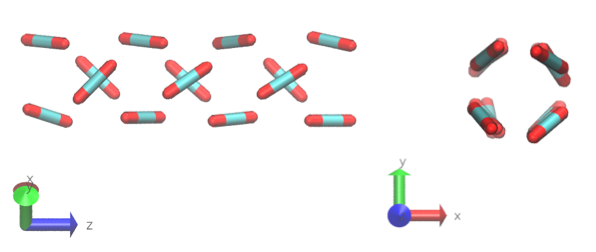} & \includegraphics[width=0.20\textwidth]{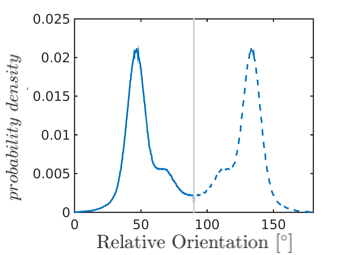} \\ 
\hline 
\rule[-1ex]{0pt}{2.5ex}  C & \includegraphics[width=0.35\textwidth]{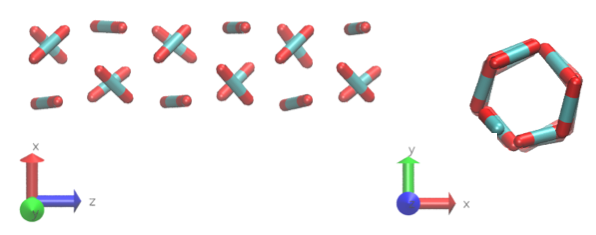} & \includegraphics[width=0.20\textwidth]{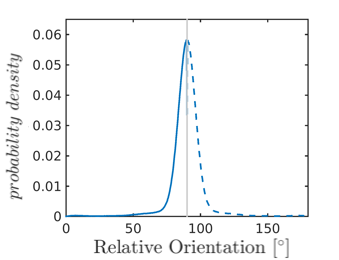} \\ 
\hline 
\rule[-1ex]{0pt}{2.5ex}  D & \includegraphics[width=0.35\textwidth]{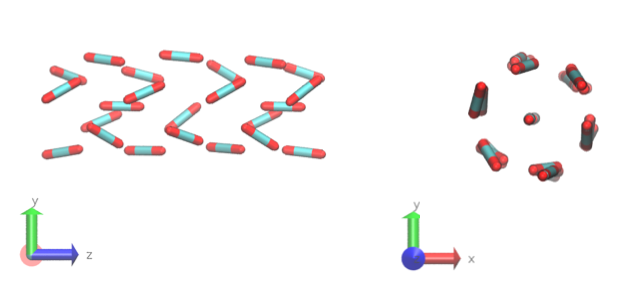} & \includegraphics[width=0.20\textwidth]{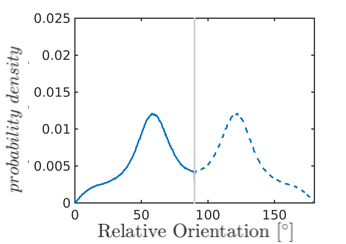} \\ 
\hline 
\hline 
\end{tabular} 
\end{center}
\label{imageordered}
\end{table*}

\begin{figure*}[t]
\centering
\includegraphics[width=0.8\textwidth]{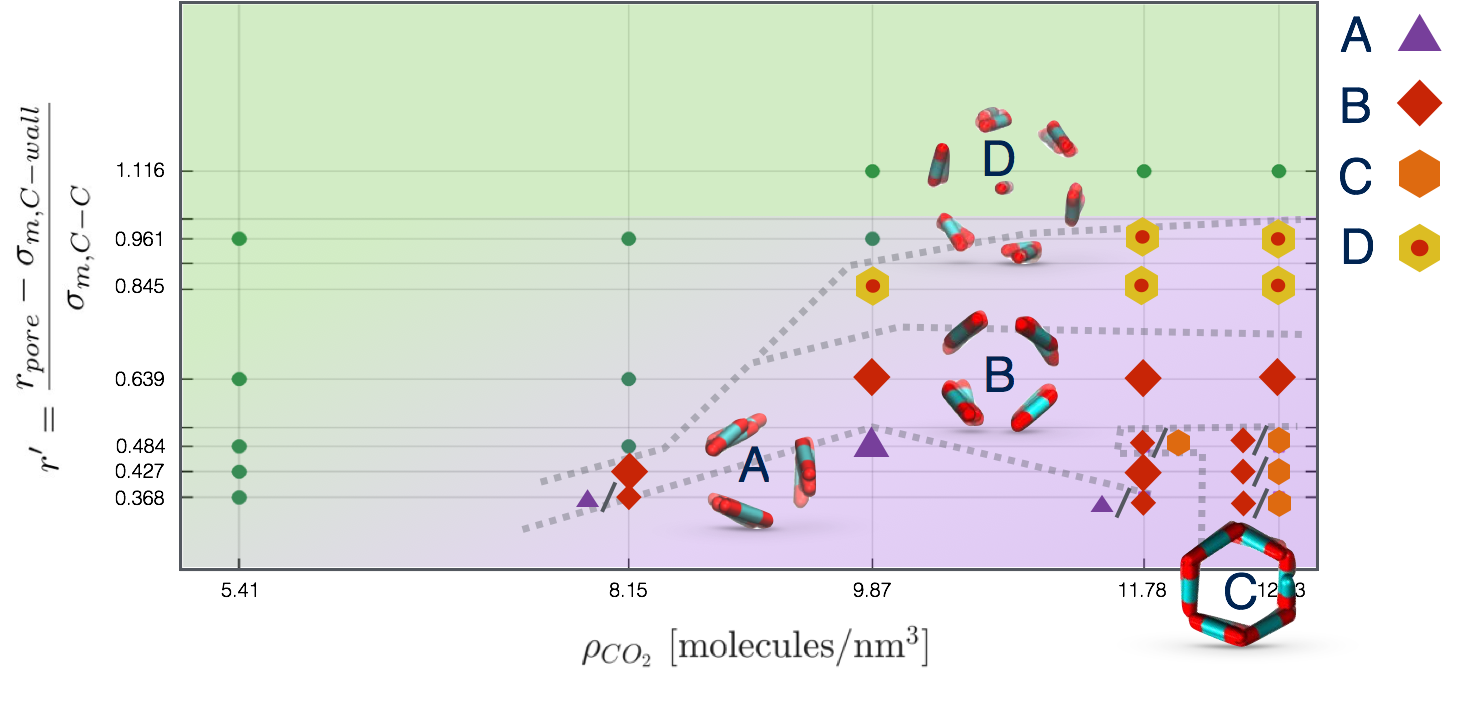}
\caption{Qualitative phase diagram of CO$_2$ confined in weakly attractive cylindrical nanopores. In green is represented the region of phase space characterised by disordered configurations, while we report in violet the region of phase space where ordered structures are found. Configurations yielding one or more amongst packing A, B, C, and D are appropriately marked. Qualitative phase boundaries have been reported in order to highlight how simulations yielding similar structures tend to cluster in parameter space, suggesting that the spontaneous ordering process observed from MD reflects the underlying thermodynamic stability of the packings.}
\label{phase_diagram}
\end{figure*}

As far as the packing of these four ordered structures is concerned, it can be seen that the characteristic angle distribution  represents a clear fingerprint of their molecular arrangement (Table~\ref{imageordered}), while the coordination number appears to be a function of the total density inside the pore and cannot clearly discriminate between different arrangements. In particular, the orientation distribution of B presents well-defined peaks in $\theta_1$ = 46.2$^\circ$ and symmetrically in $\theta_2$ = 133.8$^\circ$, while C a very sharp single one in $\theta$ = 90$^\circ$. Structure A, instead, appears to be an intermediate between B and C both from the inspection of the coordinates and from its characteristic angle distribution. Indeed, while the most densely populated orientations closely resemble those of B, the peak at 90$^\circ$ is more pronounced than in B. Form D present instead wider peaks in $\theta_1$ = 57.9$^\circ$ and the additional peak $\theta_2$ = 122.1$^\circ$.

Interestingly, three out of four ordered structures, more precisely A, B, and D, form spontaneously in unbiased MD simulations; phase C, instead, is generated from metadynamics simulations performed in conditions that spontaneously yielded other ordered arrangements  (violet area in the $r^\prime$/$\rho_{CO_2}$ phase diagram in Figures~\ref{imagegofr} and\ref{phase_diagram}); structure C is confirmed to be a metastable state for the system possessing a finite lifetime through 20 ns long unbiased MD simulations, initialised in configuration C, which do not display any sign of instability transition.
\newline It should be noted that the exploration carried out with WTMetaD yields additional ordered arrangements, which, however, do not survive unbiased MD and are therefored considered to be unstable (see SI).

In addition metadynamics further confirms that for $\rho_{CO_2}$-$r^\prime$ conditions that induce spontaneous ordering, the disordered liquid state is strongly disfavoured with respect to the ordered phases. Indeed even in extended WTMetaD simulations, characterised by the deposition of significant bias, no melting event can be observed, while transitions between ordered arrangements are routinely detected. This finding is consistent irrespective of the set of CVs used to enhance the exploration and suggests that the energetic barrier associated with melting is higher than the barriers associated with solid-solid interconversion.

Furthermore, the conversion from hollow (A, B, C) to filled (D) ordered structures and vice versa is not sampled during WTMetaD simulations;  this transition appears unlikely in the conditions investigated since, as discussed before, different values of $r^\prime$ can accommodate different numbers of CO$_2$ layers. 

On the other hand, transition between phase A, B and C is possible and consistently sampled. More precisely, it takes place though a gradual reorganisation of the layers that starts in a localised region and then spreads through the entire pore length. 
Interestingly, WTMetaD allows to sample mixed arrangements of different packings, (e.g. half ordered as B and half as C) which reveal to be stable in subsequent MD simulation, confirming that solid-solid interconversion is an activated event. Moreover, in these structures it is common to find boundary regions in which the order is not well-defined, effectively acting as defects within the pore.

%% file: tex_files/conclusion.tex
In this work we present a systematic investigation of the effects of confinement in cylindrical nanopores on CO$_2$, as a function of pore size, wall potential and CO$_2$ density.
To systematically detect the formation of ordered packings we have devised a general analysis approach based on the calculation of the time-dependent Bhattacharyya distance between a reference probability density and the instantaneous probability density of the relative orientation of nearest neighbours. Here we have adopted as a reference an ideal random packing, which closely approximates the distribution of relative orientations in the bulk of the liquid phase. This analysis approach is general and we anticipate its applicability to a much wider range of phase transitions in molecular solids. 
  
We find that cylindrical confinement induces polymorph selection by inhibiting the nucleation of known solid forms of CO$_2$ while at the same time inducing the organisation of CO$_2$ into a series of distinct ordered arrangements, which do not resemble any of the known CO$_2$ polymorphs. 
Using a combination of unbiased MD and WTMetaD we identify four ordered packing that are stable in the region of parameter space corresponding to small pore radii and large densities.  These configurations have been labelled A, B, C and D following the progression in the number of CO$_2$ molecules visible in a cross-section of a pore. Interestingly, none of these configurations corresponds to known bulk structures. It should be noted that, while A, B, and D emerge spontaneously from unbiased MD, C is found from the enhanced exploration of the configuration space achieved with WTmetaD.

In Figure \ref{phase_diagram} we report a qualitative phase diagram that summarizes the identified structures and the regions of parameter space where such structures have obtained. Markers point the location in parameter space where specific packings have been observed. Cases in which multiple configurations are indicated on the phase diagram represent points in which multiple arrangements have been observed either with unbiased MD or through WTMetaD. We observe that, in the ordered region of the phase diagram WTMetaD was quite efficient in promoting transitions between ordered packings, however it was unable to induce melting. Despite its qualitative character, this observation suggests that within the ordered region of the phase diagram solid-solid interconversion is energetically favoured with respect to liquid mediated transitions.

Finally we observe that, even in the disordered region confinement induces a short-range organisation of the liquid phase, in which the distribution of relative orientations departs from an ideal random packing, as shown in Figure \ref{imagegofr}. At large volumes and low densities the bulk behaviour is recovered as expected.   

To conclude we note how such a simple model system such as CO$_2$ confined in a simple nanometric pore model unveils remarkable structural complexity, showcasing the profound effect of confinement on the phase behaviour of molecular solids. This work paves the way for a systematic assessment of CO$_2$ packing polymorphism in systems characterised by realistic pore geometries and surface chemistry.

%% file: tex_files/acknowledgements.tex
The        authors        acknowledge        EPSRC        (Engineering        and        Physical        Sciences        Research        Council)        for        PhD        scholarship,        and   UCL       
Legion        High        Performance        Computing        Facility                for        access        to        Legion@UCL        and        associated        support        services,        in        the       
completion       of       this       work.

%% file: tex_files/SI.tex
\paragraph*{Ideal random distribution of relative orientations.} 
All possible relative orientations of two unit vectors in three dimensional space can be mapped in spherical coordinates as a function of the polar angle $\theta$ and the azimuthal angle $\phi$. 
Randomly oriented molecules are characterised by a flat probability distribution across all possible orientations. The homogeneous probability density on the surface of a sphere of unit radius has the constant value $p_{rand}(\theta,\phi)=1/4\pi$, which satisfies the normalization condition: 
\begin{equation}
\int_{\theta=0}^\pi \int_{\phi=0}^{2\pi} \frac{1}{4\pi}{d\phi}\sin\theta{d\theta}=1
\end{equation}

The angle used to map relative orientation between two neighbouring molecules in this work corresponds to the polar angle $\theta$. The probability  in $\theta$ associated with a random arrangement in spherical coordinates can thus be obtained by integrating out $\phi$ as follows: 
\begin{equation}
P_{ref}(\theta)=\int \int_{\phi=0}^{2\pi} \frac{1}{4\pi}{d\phi}\sin\theta{d\theta}=\int \frac{1}{2}\sin\theta{d\theta}
\end{equation}

The probability density in $\theta$, in this work used as a fingerprint of the molecular arrangement is then obtained by differentiating $P_{ref}(\theta)$ with respect to $\theta$:
\begin{equation}
p_{ref}(\theta)=\frac{dP_{ref}(\theta)}{d\theta}=\frac{1}{2}\sin\theta
\end{equation}

\begin{figure*}[t]
\includegraphics[width=1\textwidth]{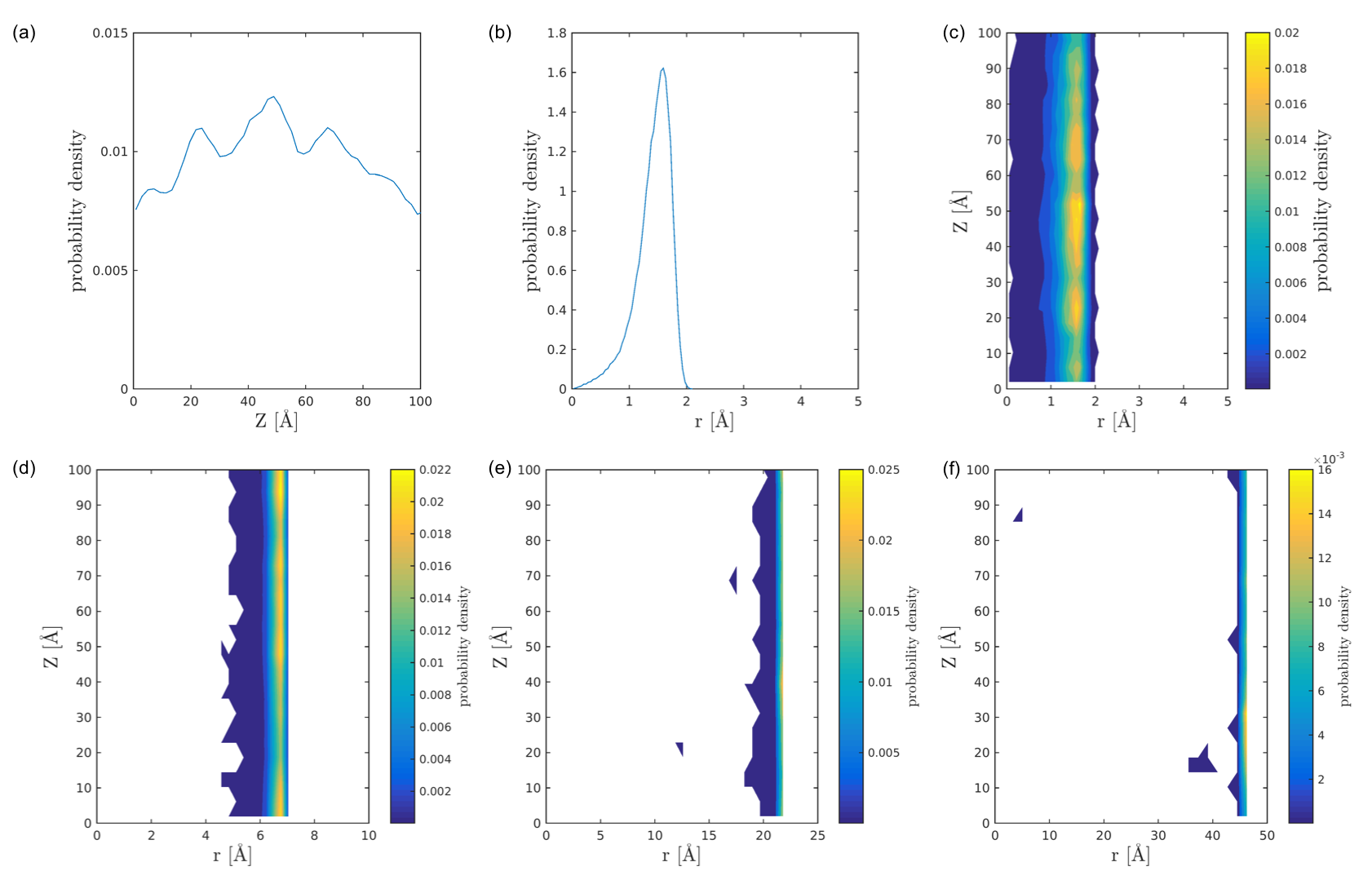}
\caption{Probability density of the location of a single CO$_2$ molecule inside pores with $\sigma_{wall}$ = 0.34 nm and diameter 1 nm (a - c), 2 nm (d), 5 nm (e) and 10 nm (f). In (a) and (b) we report the probability density as a function of only the location along the molecular axis, $z$ (a), and of only the radial position, $r$ (b), to better visualise the different role they play: negligible $z$, while determinant $r$. In all graphs, the radial position axis ranges from zero, i.e. the pore axis, to the value of the pore radius, reported in $\AA$.}
\label{image1co2}
\end{figure*}

\paragraph*{One CO$_2$ molecule.} 
Initially, we perform standard MD runs with only one particle confined in differently sized pores (d$_{pore}$ = 1, 2, 5, 10 nm, l$_z$ = 10 nm). These simulations last for about 95, 138, 125 and 30 ns, respectively and the potential of the wall is set to the reference $\sigma_{wall}$ = 0.34 nm; the initial position of CO$_2$ is random, but clearly detached from the confinement barrier. This analysis aims at giving an insight into the solely effect of the wall on the motion of a particle when the interaction with other CO$_2$ is absent.
\newline Figure~\ref{image1co2} shows that the single molecule tends to adsorb on the pore, positioning at a distance from the wall that is slightly smaller than the location of the Lennard-Jones well, i.e. $\sigma_{m,C-wall}$ (0.345 nm). Moreover, all the positions along the pore axis are equally likely, and thus the height does not play a major role in the outcome. The noise in the probability density along \textit{z} (Figure~\ref{image1co2} (a)) is probably due to limited sampling of the area, related to the computational time allowed.
\newline These simulations confirm the strong impact that the pore has on confined CO$_2$ in the radial direction, in particular the tendency of adsorbing them.

\begin{figure*}[t]
\includegraphics[width=1\textwidth]{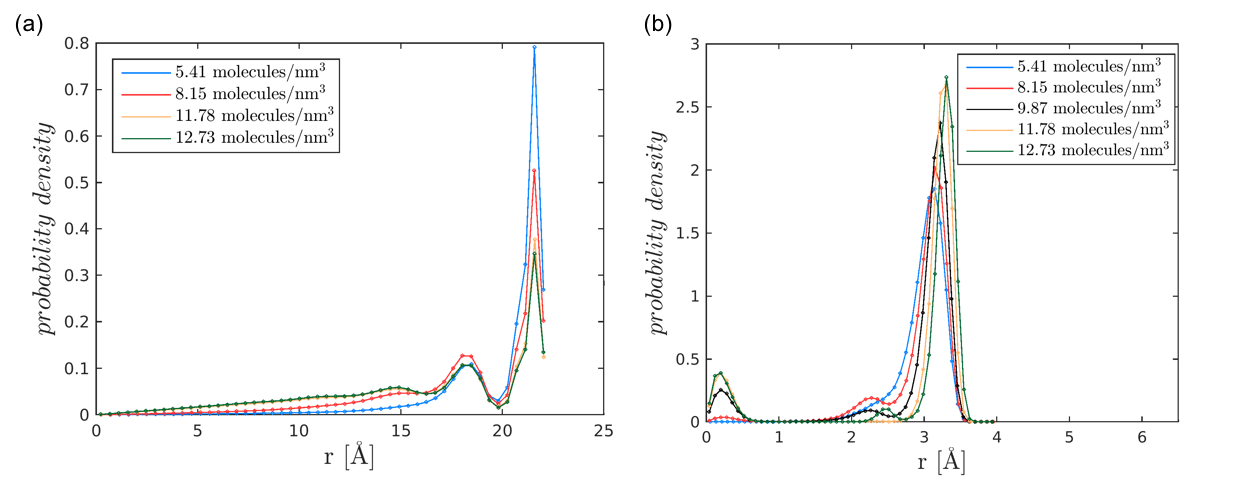}
\caption{Probability density profiles of the position of CO$_2$ molecules in the radial direction of the pore, for the range of densities investigated at different values of $r^\prime$, namely  6.847 in (a), and 0.961 in (b). Along the pore radius, zero corresponds to the cylinder axis, while the maximum value reported to the radius of the pore.}
\label{imagepeaksSI}
\end{figure*}

\paragraph*{Analysis of the radial distribution of molecules in the pore.}
We present in Figure~\ref{imagepeaksSI} the analysis of unbiased MD trajectories, focusing on the position of  CO$_2$ molecule in the radial direction of the pore, for two examples not reported in section 3, namely $r^\prime$ = 6.847 (a) and  0.961 (b). In these two cases as well, it is possible to notice that if the pore allows the formation of more layers of molecules or a bulk-like filling, the height of the first peak corresponding to the adsorbed layer decreases with growing density (a); on the other hand, if the pore has no space to accommodate more layers, the height of the peaks increases with $\rho_{CO_2}$.

\paragraph*{WTMetaD set-up.}
Well-Tempered Metadynamics simulations are performed biasing either $\lambda_I$ or $\lambda_B$. A typical simulation has initial Gaussian height equal to 4 kJ/mol, i.e. $\sim$ 1.5 kT; the width of such Gaussian depends instead on the $r^\prime$-$\rho_{CO_2}$ conditions, and spans from 3$\times$10$^{-4}$ to 0.6. A biasfactor of 20 is generally employed, but some conditions were explored also with a higher value of this parameter  (50).

\begin{figure*}[t]
\includegraphics[width=1\textwidth]{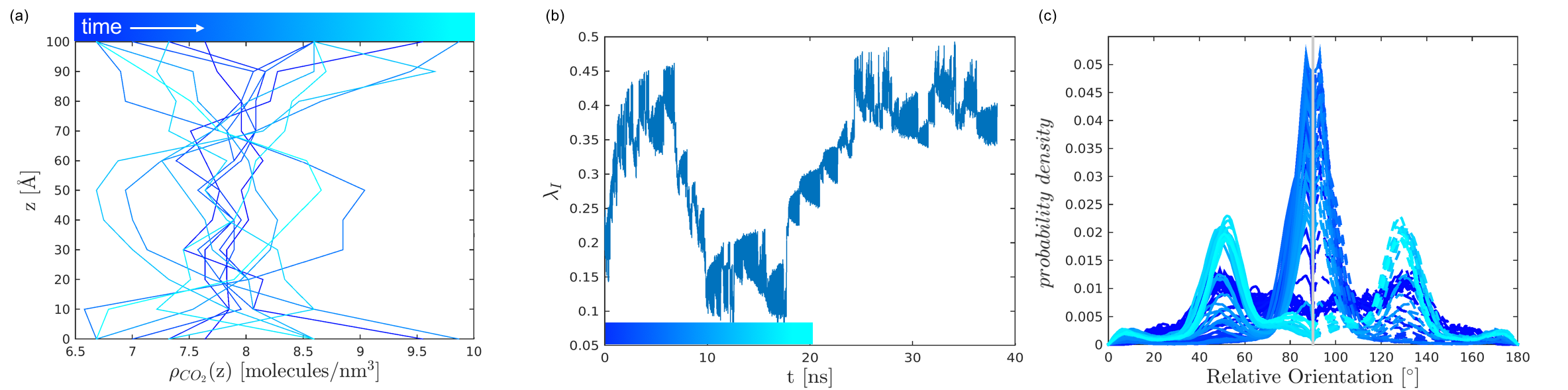}
\caption{Explorative WTMetaD biasing $\lambda_I$. (a) Axial density profile over time for the creation of a droplet at  $r^\prime$ = 2.075 - $\rho_{CO_2}$ = 8.15 molecules/nm$^3$. (b-c) Time-evolution of the CV $\lambda_I$ in (b) and of the characteristic angles over the first 20 ns in (c) at $r^\prime$ = 0.484 - $\rho_{CO_2}$ = 12.73 molecules/nm$^3$. The time scale for the orientation  distribution in (c) is highlighted in (b) by dark blue to light blue shades on the time axis. }
\label{imageWTMetaDordered}
\end{figure*}

\paragraph*{WTMetaD trajectories.} 
We hereafter present two representative examples of explorative WTMetaD simulations employing $\lambda_I$ as CV: one in case of liquid conditions (Figure~\ref{imageWTMetaDordered} (a), at $r^\prime$ = 2.075 - $\rho_{CO_2}$ = 8.15 molecules/nm$^3$), and the other for ordered structures (Figure~\ref{imageWTMetaDordered} (b-c), at $r^\prime$ = 2.075 - $\rho_{CO_2}$ = 12.73 molecules/nm$^3$). As discussed in section 3, WTMetaD did not enhance the sampling of either ordering phenomena  in $r^\prime$-$\rho_{CO_2}$ conditions identified as liquid from MD or melting events for organized structures; however, some interesting results can be observed. In the case of liquid state, biasing $\lambda_I$ leads to the creation of an unstable droplet with higher density than the nominal one, that migrates along the $z$-axis, as shown in Figure~\ref{imageWTMetaDordered} (a).
\newline On the other hand, in WTMetaD simulations in ordered areas of the $r^\prime$-$\rho_{CO_2}$ phase diagram, despite not observing melting, many transitions between ordered structures take place, as reported in Figure~\ref{imageWTMetaDordered} (b-c). In these explicative plots, we present a B-C conversion, where B is characterised by low value of $\lambda_I$ and a 2-peak angle distribution, while C has higher $\lambda_I$ and a sharp peak on 90$^\circ$ in its angle distribution. Interestingly, it is also possible to notice that during a transition, $\lambda_I$ changes gradually: indeed, as we highlighted in the results, the interconversion between ordered structures generally takes place by progressive rearrangement of unit cells along the $z$-axis.

\begin{figure*}[t]
\includegraphics[width=0.65\textwidth]{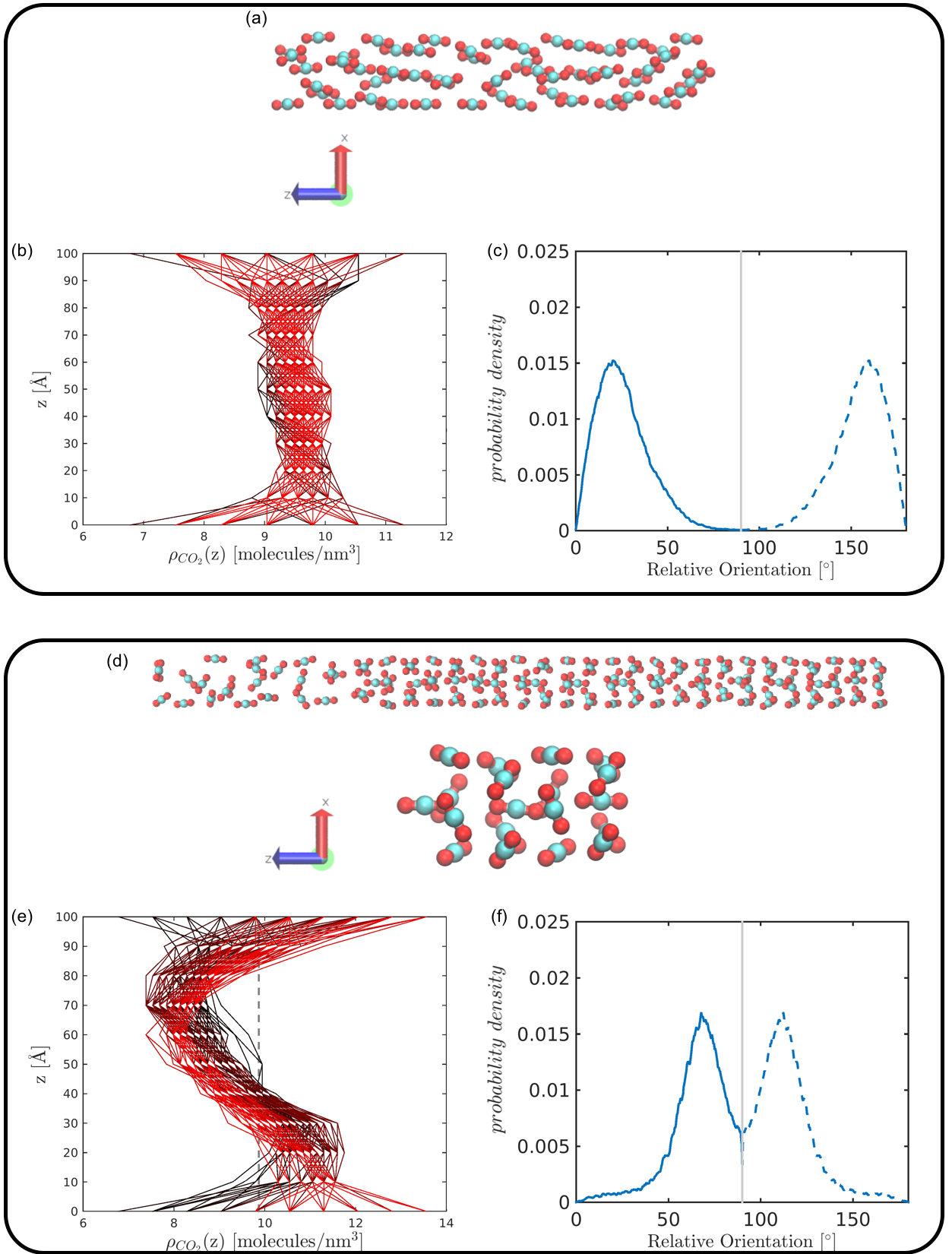}
\caption{Examples of unstable ordered structures emerging from WTMetaD with $\lambda_I$ at $r^\prime$ = 0.961 - $\rho_{CO_2}$ =  9.87 molecules/nm$^3$. In particular, (a) to (c) report snapshots, density profile in the pore over WTMetaD time (shading from black to red for growing time), and angle distribution for the first arrangement presented; (d) to (f), instead, present the same analysis for the second one.}
\label{imageunstable}
\end{figure*}

\paragraph*{Unstable ordered structures.} 
Thanks to both MD and WTMetaD, we identify four stable ordered structures. However, WTMetaD simulations explore a bigger number of configurations that instead are not stable in unbiased simulations, and reorganise in one of the main four. Two significant examples are reported in Figure~\ref{imageunstable}. The first of these two structures (Figure~\ref{imageunstable} (a-c)) has  CO$_2$ molecules parallel to each other, aligned along the  $z$-axis, and the density is the same throughout the pore length;  the second arrangement presented  in Figure~\ref{imageunstable} (d-f) has well-defined characteristic angles, similar to bulk phase I, as well as a more compact packing than the other phases; such compact packing is achieved by locally increasing the molecular density, which results in areas of the pore with $\rho_{CO_2}$ below the nominal value.
